%% file: metal_rich_fuel.tex
\documentclass[trackchanges,twocolumn]{aastex701}

\input{defs.tex}

\usepackage{booktabs} 
\usepackage{hyperref}
\usepackage{natbib}

\defcitealias{Choi2024}{Paper~I} 
\usepackage[dvipsnames]{xcolor}

\usepackage{xcolor}

\begin{document}
\title{Recycled Gas Dominates the Metal-rich Fuel of Supermassive Black Holes}
\shorttitle{Metal-rich Fuel for SMBHs}
\shortauthors{Kwak et al.}
    
\author[orcid=0009-0006-8864-6472,sname='Kwak']{Dongyun Kwak}
\affiliation{Department of Physics, University of Seoul, 163 Seoulsiripdae-ro, Dongdaemun-gu, Seoul, 02504, Republic of Korea}
\email{dongyunkwak@uos.ac.kr}

\correspondingauthor{Ena Choi}
\author[orcid=0000-0002-8131-6378,gname='Ena',sname='Choi']{Ena Choi} 
\affiliation{Department of Physics, University of Seoul, 163 Seoulsiripdae-ro, Dongdaemun-gu, Seoul, 02504, Republic of Korea}
\email[show]{enachoi@uos.ac.kr}

\author[orcid=0000-0001-6470-7476,gname='Hannah',sname='Jhee']{Hannah Jhee}
\affiliation{Department of Physics, University of Seoul, 163 Seoulsiripdae-ro, Dongdaemun-gu, Seoul, 02504, Republic of Korea}
\email{hannah.jhee@uos.ac.kr}

\author[orcid=0000-0002-6748-6821,sname='Somerville']{Rachel S. Somerville}
\affiliation{Center for Computational Astrophysics, Flatiron Institute, 162 5th Ave., New York, NY 10010, USA}
\email{rsomerville@flatironinstitute.org}

\author[orcid=0000-0002-7314-2558,sname='Naab']{Thorsten Naab}
\affiliation{Max-Planck-Institut für Astrophysik, Karl-Schwarzschild-Straße 1, 85741 Garching, Germany
}
\email{naab@MPA-Garching.MPG.DE}

\author[orcid=0000-0002-3301-3321,sname='Hirschmann']{Michaela Hirschmann}
\affiliation{Institute for Physics, Laboratory for Galaxy Evolution and Spectral modelling, \\ Ecole Polytechnique Fédérale de Lausanne, Observatoire de Sauverny, Chemin Pegasi 51, 1290 Versoix, Switzerland 
}
\email{michaela.hirschmann@epfl.ch}

\author[orcid=0000-0001-6363-8069,sname="Shin"]{Jaejin Shin}
\affiliation{Department of Astronomy and Space Science, Sejong University, 209 Neungdong-ro, Gwangjin-Gu, Seoul 05006, Republic of Korea}
\email{shin.astro@gmail.com}

\author[orcid=0000-0002-8055-5465,sname='Woo']{Jong-Hak Woo}
\affiliation{Astronomy Program, Department of Physics and Astronomy, Seoul National University, Seoul 08826, Republic of Korea
}
\email{jhwoo@snu.ac.kr}

\begin{abstract}
Understanding the origin and chemical properties of gas accreted by supermassive black holes (SMBHs) is essential for linking black hole growth to galaxy evolution. Using a suite of 30 high-resolution cosmological zoom-in simulations, we investigate the chemical properties of gas accreted onto SMBHs in massive galaxies with stellar masses of $10^{10.9-11.9}\,\rm M_\odot$ and black hole masses of $10^{8.5-9.7}\,\rm M_\odot$ at $z=0$. By tracing the full cosmological histories of individual gas particles, we identify their origins and enrichment pathways. The accreted gas is classified into four categories: ``early'' gas accreted during the early assembly phase of the main halo, ``external'' gas originating from other galaxies or subhalos, ``recycled'' gas enriched through stellar evolution processes within the primary galaxy, including asymptotic giant branch (AGB) winds and supernova ejecta, and ``smooth'' gas accreted from the intergalactic medium.
We find that recycled gas dominates the accretion budget and is already metal rich at early epochs. Gas from other origins typically undergoes gradual chemical enrichment within the galactic environment prior to black hole accretion. The mean abundance ratios show only weak redshift evolution and are broadly compatible with the high metallicities inferred for quasar broad-line regions. Our results suggest that metal-rich gas supply to SMBHs arises naturally from cosmological galaxy evolution and stellar recycling.
\end{abstract}


\section{Introduction}
Over the past several decades, both observations and cosmological simulations have demonstrated that supermassive black holes (SMBHs) and their host galaxies evolve in close connection \citep[e.g.][]{Silk1998,Richstone1998,Heckman2014,Sijacki2015}. 
Tight empirical correlations between SMBH mass and host-galaxy properties, such as bulge mass, stellar velocity dispersion, and total binding energy, suggest a physical link between SMBH growth and galaxy assembly \citep{Magorrian1998,Kormendy2013,McConnell2013,Woo2013,Woo2015}. Within the hierarchical framework of galaxy formation, gas accretion, star formation, and feedback processes collectively drive this coevolution \citep[e.g.][]{Somerville2015,2017ARA&amp;A..55...59N}. In particular, energetic feedback from accreting black holes (BHs), commonly referred to as active galactic nucleus (AGN) feedback, plays a central role in regulating both star formation and SMBH growth by heating, expelling, or redistributing interstellar gas \citep{Springel2005a,Sijacki2007,Somerville2008a,Choi2012a,Dubois2012}.

While the coevolution of SMBHs and galaxies has been widely investigated \citep[e.g.][]{Terrazas2020,Habouzit2021,Iyer2025}, the physical and chemical nature of the gas that fuels SMBH growth has not yet been explored in detail. In particular, the chemical composition of gas accreted onto SMBHs, which encodes the integrated history of star formation, stellar mass loss, and metal recycling within galaxies, remains poorly characterized in cosmological simulations. Despite extensive observational studies of quasar metallicities, the chemical evolution of gas accreted onto SMBHs has remained largely unexplored in a fully cosmological context.

Observationally, the chemical properties of gas in the immediate vicinity of luminous quasars have been extensively studied through emission-line diagnostics of the broad-line region (BLR) \citep{Barth2003,Maiolino2003}. These studies have consistently found that the BLR in the most luminous quasars exhibits supersolar metallicities \citep{Hamann1992,Dietrich2003,Nagao2006}, even at very high redshift ($z>6$; \citealt{Kurk2007,Jiang2007,Juarez2009,Mortlock2011,Onoue2020,Wang2022}). Moreover, many surveys report little or no redshift evolution in inferred metallicity over cosmic time \citep{Mazzucchelli2017,Shin2021,Sameshima2017,Sameshima2020,Yoshii2022,Jiang2024}. 
The physical origin of this apparent lack of evolution, however, remains unclear, particularly given the dramatic changes in galaxy growth, star formation, and chemical enrichment across cosmic history.

Several studies have also suggested that quasar metallicity indicators may be linked to nuclear-scale physical conditions associated with BH accretion \citep{Shemmer2004,Mayer2015}. In particular, observed correlations between BLR metallicity indicators and Eddington ratio imply that the chemical properties of quasar environments may be connected to central accretion activity \citep[e.g.][]{Shin2013,Shin2019,Shin2021}. However, these correlations alone do not explain why quasar environments remain highly metal enriched over a wide range of cosmic time.

One possibility is that the metal-rich nature of quasar environments reflects the cosmological assembly of galaxies and the recycling of chemically enriched stellar ejecta. In galaxies dominated by old stellar populations, gas recycling from stellar mass loss is inevitable \citep[e.g.][]{Leitner2011}. Hydrodynamic simulations of an isolated elliptical galaxy have shown that recycled gas alone can fuel episodic AGN activity and central starbursts even in the absence of galaxy mergers and cosmological cooling of gas \citep{Ciotti2007}. These results suggest that stellar mass loss may provide a persistent and metal-rich gas reservoir capable of sustaining BH accretion over long timescales.

Testing this picture requires tracking not only how gas reaches SMBHs, but also where it originates and how it becomes chemically enriched in a cosmological context. Particle-tracing techniques applied to cosmological zoom-in simulations provide a powerful tool for this purpose, as they allow the full dynamical and chemical histories of individual gas particles to be followed. Such methods have been used to study the origins of stars \citep{Oser2010,Shipp2023,Gandhi2024}, CGM gas \citep{Hafen2019,Hafen2020,Rottgers2020}, galactic winds \citep{Brennan2018,Choi2020,Mercedes-Feliz2023,Mercedes-Feliz2025}, and the baryon cycle in galaxies \citep{Christensen2016,Daniel2017a,Ho2019,Tollet2019,Mitchell2020}. However, relatively few studies have applied particle tracing specifically to gas accreted by SMBHs.

Among these, \cite{Bellovary2013} traced gas particles accreted onto intermediate-mass BHs in high-redshift galaxies ($z\sim4$) and found that BH growth largely reflects the global gas composition of the host galaxy, with cold inflows dominating the accreted mass. Extending this approach to a Milky Way–mass galaxy at $z=0$ with a merger-rich assembly history, \cite{Sanchez2018} showed that SMBHs preferentially accrete low-angular-momentum gas, emphasizing that the angular momentum of gas upon halo entry is more important than its accretion mode (e.g., merger-driven versus smooth accretion) in determining BH fueling. While these studies provided valuable insights into the dynamical origins of accreted gas, they did not examine the chemical enrichment histories or metallicity of the fueling material.

In this work, we investigate the origins and chemical enrichment pathways of gas accreted onto SMBHs in a suite of 30 high-resolution cosmological zoom-in simulations of massive galaxies at the group scale, originally introduced in \citet{Choi2017}. Our sample spans a wide range of SMBH masses (about 1.2 dex at $z\sim0$) and galaxy assembly histories, allowing us to explore how BH fueling depends on environment and evolutionary stage. By tracing the full trajectories of individual gas particles, we classify accreted material into four origin categories---Recycled, Early, External, and Smooth---and analyze how their distinct dynamical histories shape their chemical properties.

In massive galaxies dominated by old stellar populations, recycled gas from stellar mass loss is expected to be a major contributor to the available gas reservoir \citep{Leitner2011,Segers2016,Eisenreich2017, Pellegrini2018}. Building on this expectation, our analysis explicitly tracks recycled and externally supplied gas to quantify their relative contributions to BH fueling and to characterize their enrichment histories in a fully cosmological context. This approach allows us to determine whether the metal-rich nature of gas fueling SMBHs arises primarily from internal stellar recycling, external accretion, or cumulative chemical pre-processing within the galactic environment.

Our study builds on the particle-tracing methodology developed in \cite{Choi2024}, hereafter \citetalias{Choi2024}, which demonstrated that recycled stellar material dominates the mass budget of gas accreted by SMBHs across cosmic time. While that work focused primarily on the dynamical origins and relative contributions of different gas components in a smaller sample, we extend the analysis here to a larger suite of 30 galaxies and to the chemical domain by examining elemental abundance ratios such as $\mathrm{[Fe/H]}$ and $\mathrm{[Mg/Fe]}$. This allows us to directly link the dominant role of recycled gas to the metallicity of BH fueling material and to explore how chemical enrichment proceeds along different accretion pathways in a fully cosmological context.

The paper is organized as follows. Section~\ref{sec:method} describes the simulations and particle-tracing methods used to identify accreted gas particles and classify their origins. 
Section~\ref{sec:Result} presents the main results on the chemical composition, enrichment histories, and dynamical properties of the accreted gas. In Section~\ref{sec:discussion}, we discuss the physical implications of our findings in the context of cosmological galaxy evolution and observational constraints on quasar metallicities. Finally, Section~\ref{sec:summary} summarizes our main conclusions.

\begin{figure*}
\centering
\makebox[\textwidth][c]{%
    \includegraphics[width=1.0\textwidth]{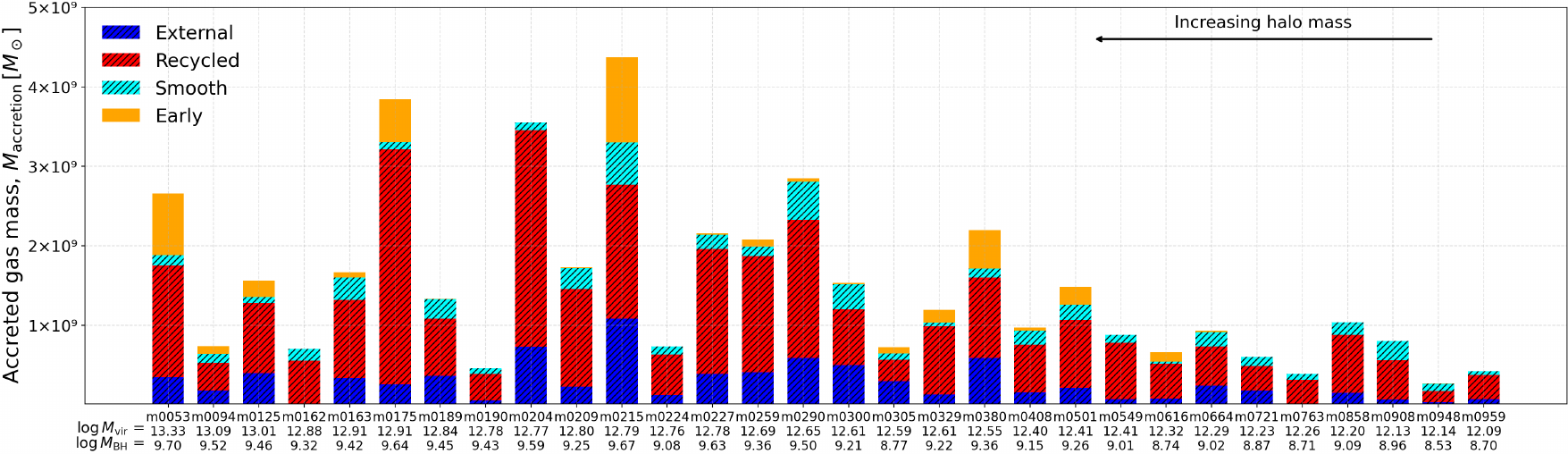}%
}
\caption{
Census of the cosmological origin of gas particles accreted onto the central SMBHs in each simulated galaxy. The accreted gas is classified into four categories: Early (orange), Smooth (cyan), Recycled (red), and External (blue). The first row beneath the x-axis lists the virial mass of each galaxy, while the second row indicates the final BH mass ($M_{\rm BH}$) at $z=0$. Across the sample, gas originating from stellar mass loss within the main galaxy (Recycled) constitutes the largest fraction of the accreted gas budget, whereas the relative contributions from Early, Smooth, and External gas vary significantly from system to system. Here, the  mass budget includes only gas accreted onto the SMBHs and excludes mass growth through BH--BH mergers.}
\label{fig:census}
\end{figure*}

\section{Simulation and Method}\label{sec:method}
\subsection{Simulation}\label{subsec:simulation}
We analyze a subset of the cosmological hydrodynamical zoom-in simulations of massive galaxy and SMBH formation originally presented in \citet{Choi2017}. Here we summarize the key aspects of the numerical setup relevant for this study, with particular emphasis on the treatment of BH growth and chemical enrichment. A detailed description of the full simulation suite can be found in \citet{Choi2017}.

The zoom-in initial conditions are based on the massive halo sample of \citet{Oser2010}, selected from a dark-matter-only simulation with a comoving box size of $L=72$~Mpc and WMAP3 cosmological parameters \citep[$h=0.72$, $\Omega_{\mathrm{b}}=0.044$, $\Omega_{\mathrm{dm}}=0.26$, $\Omega_{\Lambda}=0.74$, $\sigma_8=0.77$, $n_s=0.95$;][]{2007ApJS..170..377S}. For each target halo, all particles within $2 \times r_{\rm vir}$ at $z=0$ are traced back to the initial redshift and replaced with high-resolution dark matter and baryonic particles. This procedure ensures that the full cosmological assembly history of each massive galaxy is resolved at high resolution.

The simulations are performed using SPHGal \citep{2014MNRAS.443.1173H}, a modified version of the GADGET-3 smoothed particle hydrodynamics code \citep{2005MNRAS.364.1105S}. To improve fluid mixing and shock capturing, SPHGal adopts a pressure--entropy formulation \citep{2001MNRAS.323..743R}, an improved artificial viscosity scheme \citep{2010MNRAS.408..669C}, and an artificial thermal conductivity \citep{2012MNRAS.422.3037R}. The particle masses are $m_{\rm dm}=2.5\times10^{7}\,\Msunh$ for dark matter and $m_{\rm gas,*}=4.2\times10^{6}\,\Msunh$ for gas and star particles. The comoving gravitational softening lengths are $\epsilon_{\rm dm}=0.89\,\kpch$ and $\epsilon_{\rm gas,*}=0.4\,\kpch$.

Star formation and chemical enrichment follow the model of \citet{2013MNRAS.434.3142A}. The simulations explicitly track the abundances of 11 chemical species (H, He, C, N, O, Ne, Mg, Si, S, Ca, and Fe) in both gas and star particles. Metal enrichment from Type~Ia supernovae, Type~II supernovae, and asymptotic giant branch (AGB) stars is included using stellar yield tables from \citet{1999ApJS..125..439I}, \citet{1995ApJS..101..181W}, and \citet{2010MNRAS.403.1413K}, respectively. With the adopted initial mass function \citep{2001MNRAS.322..231K}, approximately 30\% of the stellar mass formed is returned to the interstellar medium over $\sim$13~Gyr through stellar winds and supernova ejecta, providing a continuous source of chemically enriched recycled gas.

In our adopted chemical enrichment model, magnesium is produced predominantly by Type~II supernovae on short timescales, whereas iron receives substantial delayed contributions from Type~Ia supernovae in addition to Type~II supernovae. AGB stars primarily contribute recycled gas mass and lighter elements such as C and N through stellar mass loss. Consequently, the evolution of gas-phase $\mathrm{[Mg/Fe]}$ in the simulations reflects the relative timing of prompt and delayed enrichment channels.

Stellar ejecta are deposited onto neighboring gas particles, increasing their mass and metallicity over time. As a result, gas particles may grow more massive than the initial baryonic resolution, while stellar particles lose mass. When a gas particle exceeds twice the original mass resolution, it is split into two equal-mass particles that inherit identical physical and chemical properties, ensuring a stable numerical treatment of ongoing mass and metal injection.

Stellar feedback is implemented using the multi-channel kinetic feedback model of \citet{2017ApJ...836..204N}, which includes momentum and energy input from massive star winds, photoionization heating in \strom spheres, three-phase supernova remnants (from both Type~Ia and Type~II events), and outflows from low-mass AGB stars. Metal diffusion between gas particles is modeled following \citet{2013MNRAS.434.3142A}, allowing metals to mix through turbulent diffusion and promoting a more realistic spatial distribution of chemical elements.

BH particles with an initial mass of $10^{5}\,\Msunh$ are seeded at the centers of newly formed dark matter halos with masses above $10^{11}\,\Msunh$. BHs grow through mergers with other BHs and by accreting gas from their surroundings. Mergers are allowed only when two BHs fall within their respective SPH smoothing lengths and have relative velocities below the local sound speed. Gas accretion follows a Bondi--Hoyle--Lyttleton prescription \citep{1939PCPS...34..405H,1944MNRAS.104..273B,1952MNRAS.112..195B}, with a probabilistic soft-Bondi criterion that limits accretion to gas located within the Bondi radius \citep{Choi2012a}.

The accretion probability is further regulated by accounting for the smoothing length of gas particles and the local free-fall timescale. The accretion rate is not artificially capped at the Eddington limit. Instead, the model includes the effect of Eddington radiation pressure acting on electrons, allowing for occasional super-Eddington accretion episodes that are rapidly self-regulated by feedback processes.

Mechanical AGN feedback injects mass, momentum, and energy into the surrounding gas \citep{Choi2012a,Choi2014}, driving powerful outflows that suppress star formation and regulate BH growth \citep{Ostriker2010a,Choi2015a}. This feedback model is motivated by observations of strong AGN-driven winds \citep[e.g.,][]{Arav2020} and theoretical models of radiatively efficient accretion \citep[e.g.,][]{2000ApJ...543..686P,2004ApJ...616..688P}. To ensure accurate shock propagation, a time-step limiter is applied so that neighboring particles evolve on similar time steps.

Radiative feedback from the BH is also included in the form of Compton and photoionization heating, along with associated radiation pressure from moderately hard X-ray photons ($\sim50$~keV), following \citet{2004MNRAS.347..144S,2005MNRAS.358..168S}. Together, the mechanical and radiative feedback prescriptions effectively quench star formation in massive galaxies \citep{Choi2015a,Choi2017}. The simulated galaxies are calibrated to reproduce the local $M_{\rm BH}$--$\sigma$ relation and the stellar-to-halo mass relation.

\subsection{Gas Tracing Method}\label{subsec:swallowed gas}
To identify the cosmological origin and chemical pre-processing of gas that ultimately fuels SMBHs, we employ a particle-based tracing analysis that follows individual gas elements across cosmic time.

In this study, we trace the positions, kinematics, and chemical histories of gas particles that are ultimately accreted onto the central SMBHs in a suite of cosmological zoom-in simulations. For each galaxy, the tracing begins at the snapshot in which the BH is seeded and continues until the gas particle is accreted, allowing us to reconstruct the full evolutionary pathway of the accreted gas particle.

We first identify the virial radius ($r_{\mathrm{vir}}$) and the center of each zoom-in region using the galaxy center-finding tool \textit{Gtrace}. Gas particles accreted onto the central BH are then selected from the main galaxy, defined as the region within $r_{10} = 0.1  \times r_{\mathrm{vir}}$. For each accreted gas particle, we record its unique particle ID, three-dimensional position, velocity, and mass at every simulation snapshot. Dark matter halos and subhalos are identified using the ROCKSTAR halo finder \citep{behroozi2013a}, which employs a six-dimensional phase-space friends-of-friends algorithm to robustly track substructure across time. Following the classification scheme introduced in Figure~1 of \citetalias{Choi2024}, we classify the accreted gas particles into four categories based on their formation sites and dynamical histories.
\begin{enumerate}
    \item \textbf{Recycled gas} \\
    Defined as gas originating from stellar evolution within the main galaxy, including AGB winds and SN explosions (both Type II and Type~Ia).

    \item \textbf{Early gas} \\
    Because the SMBH formation time differs from galaxy to galaxy, identifying this category required a quantitative definition applicable across the full sample. We therefore define ``Early'' gas as particles whose first accretion onto the main halo occurs at redshift $z \geq 3$. We emphasize that this definition is introduced for classification purposes only. Recycled gas may also be produced at $z>3$ through stellar evolution within the main progenitor, but such particles are included in the Early category to avoid ambiguities associated with halo identification during the earliest stages of halo assembly.
    
    \item \textbf{External gas} \\
    Defined as gas that belonged to an external halo or subhalo at least once during its trajectory. This gas is primarily supplied through galaxy mergers, and may also include recycled gas originating from stellar evolution in other galaxies.
    \item \textbf{Smooth gas} \\
     Defined as the remaining particles that do not satisfy the Recycled, Early, or External criteria, including gas not initially bound to any halo but later accreted onto the main halo.
\end{enumerate}

By tracing individual gas particles in this manner, our analysis naturally captures both large-scale environmental effects associated with massive galaxies and internal processes such as star formation, stellar mass loss, and gas recycling.

\section{Results} \label{sec:Result}

\begin{figure*}
\makebox[\textwidth][c]{%
    \includegraphics[width=0.51\textwidth]{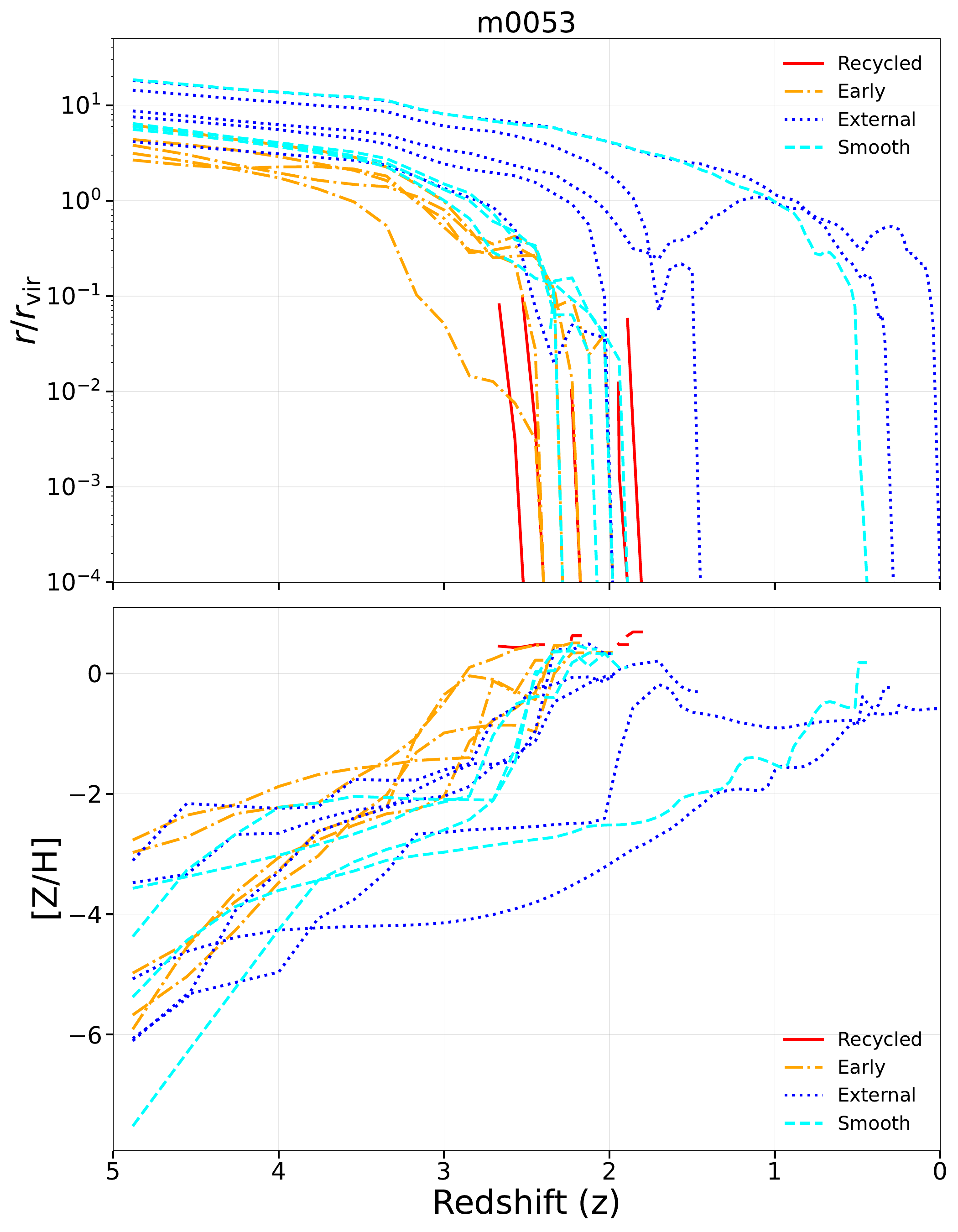}%
    \includegraphics[width=0.51\textwidth]{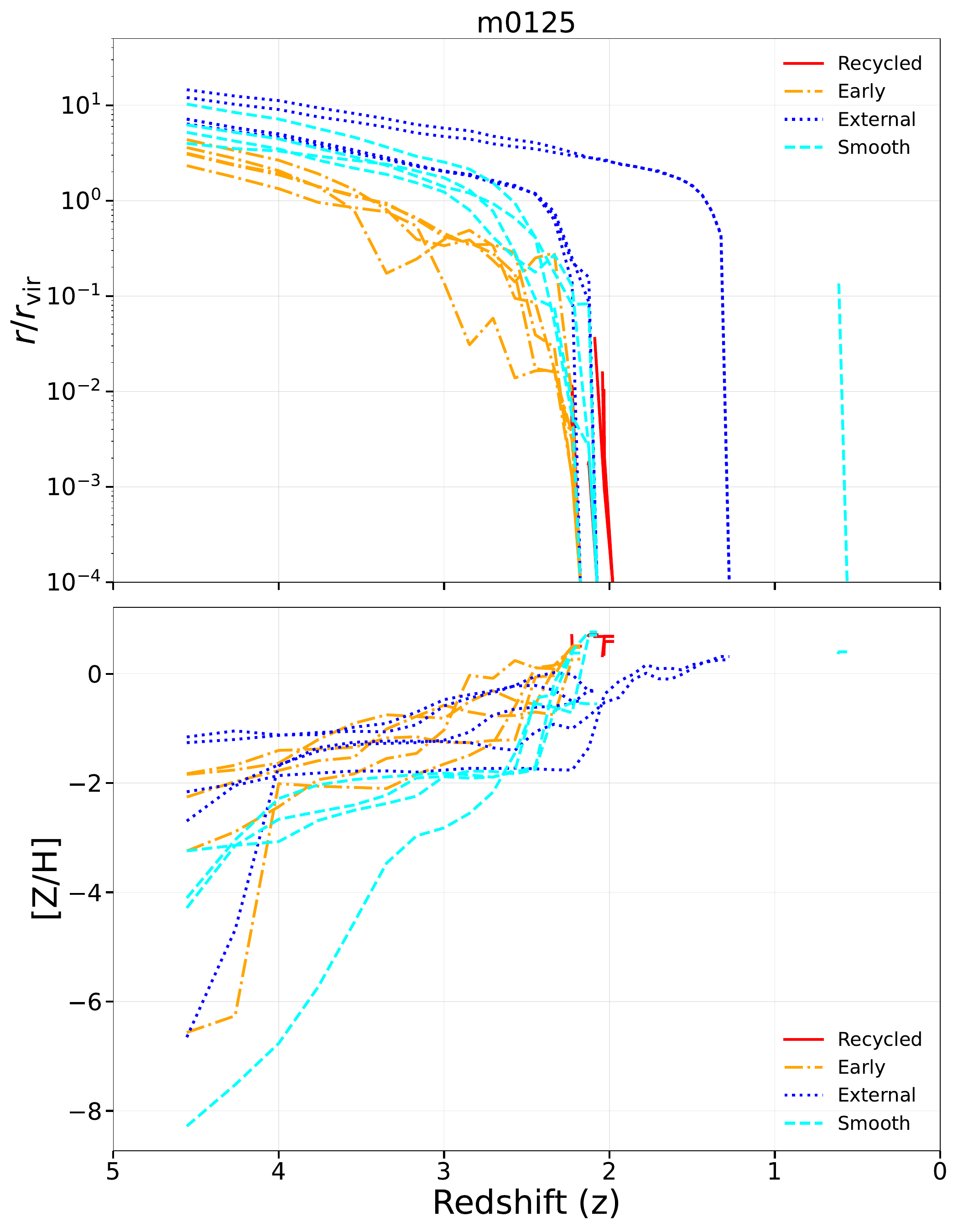}%
}

\caption{Representative evolutionary pathways of gas particles accreted onto the central SMBHs in two example galaxies, m0053 (left) and m0125 (right).
For each gas-origin category, five representative particles are shown.
{\it(Top)} Radial trajectories of the accreted gas particles, expressed as distances from the galaxy center normalized by the virial radius ($r/r_{\mathrm{vir}}$), illustrating their large-scale dynamical histories prior to being accreted.
Recycled, Early, External, and Smooth gas particles are shown in red, orange, blue, and sky blue, respectively.
By construction, recycled gas forms within the central region ($r_{10} = 0.1 \times r_{\mathrm{vir}}$), whereas gas from the other origin categories is initially located at larger radii and subsequently migrates inward along distinct orbital paths.
{\it(Bottom)} Redshift evolution of the gas-phase metallicity ([Z/H]) for the same particles shown in the top panels.
The metallicity histories reflect progressive chemical enrichment through stellar evolution, metal mixing, and interactions with the surrounding galactic environment.
Recycled gas is already metal-rich at formation, while gas from other origins undergoes gradual enrichment during its cosmological inflow toward the central region.}
\label{fig:trajectory}
\end{figure*}

We quantify the mass census of gas particles accreted onto the central SMBHs across the full sample of simulated galaxies. As shown in \autoref{fig:census}, gas originating from stellar mass loss within the main galaxy (Recycled) constitutes the largest fraction of the accreted gas budget in most systems. When averaged over the sample, the relative contributions of each origin to the total accreted gas mass are 58.70\% (Recycled), 19.76\% (External), 11.76\% (Smooth), and 9.77\% (Early).

This distribution indicates that gas processed through internal stellar evolution, such as supernova ejecta and AGB winds, contributes more significantly to the population of gas reaching the central SMBH than gas associated with external halos or merger-driven delivery. In this sense, the cosmological gas supply to SMBHs in massive galaxies is dominated by recycled stellar ejecta, with substantial system-to-system variation in the relative importance of other channels

This result also suggests that mergers are not the dominant source of the long-term gas supply sustaining SMBH growth in these massive galaxies. This is consistent with previous studies using the same simulation suite, which found that the AGN--merger connection is generally weak or limited to particular conditions, such as gas-rich mergers or systems with depleted internal gas reservoirs at lower redshift \citep{Sharma2024,Jhee2026}.

\autoref{fig:trajectory} illustrates representative dynamical and chemical evolutionary pathways of gas particles that are ultimately accreted onto the central SMBHs. For two example galaxies, m0053 and m0125, we show five representative particles from each gas-origin category, tracking their radial distances from the galaxy center and their gas-phase metallicities as a function of redshift. Throughout this analysis, metallicities are expressed relative to the solar abundance scale of \citet{Asplund2009}.

Recycled gas originates from stellar mass loss within the central region of the main galaxy and is therefore already chemically enriched when it is released into the interstellar medium. As a result, recycled particles exhibit high metallicities throughout their evolution, with relatively little additional enrichment prior to being accreted.

Gas classified as Early enters the main halo at $z \geq 3$, during the initial phases of halo assembly when star formation and chemical enrichment are rapidly progressing. These particles display well-defined metallicity evolution, reflecting the build-up of metals in the galactic environment at high redshift as successive generations of stars enrich the surrounding gas.

External gas originates in satellite halos or subhalos that later merge with the main system. Prior to merging, these particles orbit within their host halos and undergo chemical enrichment through local star formation and feedback processes. Following halo mergers, they are incorporated into the main halo, contributing both mass and metals to the central gas reservoir.

Smooth gas consists of material that is not bound to any halo for a significant fraction of its history. Such gas may reside in the intergalactic medium or form in the outskirts of the main halo, and subsequently migrates inward as it loses angular momentum. During this process, smooth gas can mix with metal-enriched diffuse gas expelled by stellar evolution, leading to a gradual increase in metallicity prior to accretion. 

By definition, this category excludes gas released by stellar mass loss within the central region ($r < r_{10}$), although it may include material enriched by stellar evolution at larger radii.

In the left panel of \autoref{fig:trajectory}, some external and smooth gas particles in galaxy m0053 exhibit non-monotonic radial trajectories at low redshift, characterized by an initial inward migration followed by a temporary outward displacement before eventual accretion. This behavior is likely associated with shock heating in the hot, massive halo of the elliptical galaxy ($\log M_{\rm vir} \sim 13.3$). Correspondingly, these particles can show transient decreases in [Z/H] when displaced to larger radii, where lower-metallicity gas dominates. Such rebound motions and temporary metallicity dilution highlight the coupling between gas dynamics, shock heating, and chemical mixing in the hot halos of massive galaxies.

\begin{figure*}[t]
\centering
\vspace{0pt}
\begin{minipage}[t]{0.5\textwidth}
  \centering
  \includegraphics[width=\linewidth,trim=0 0 0 0,clip]{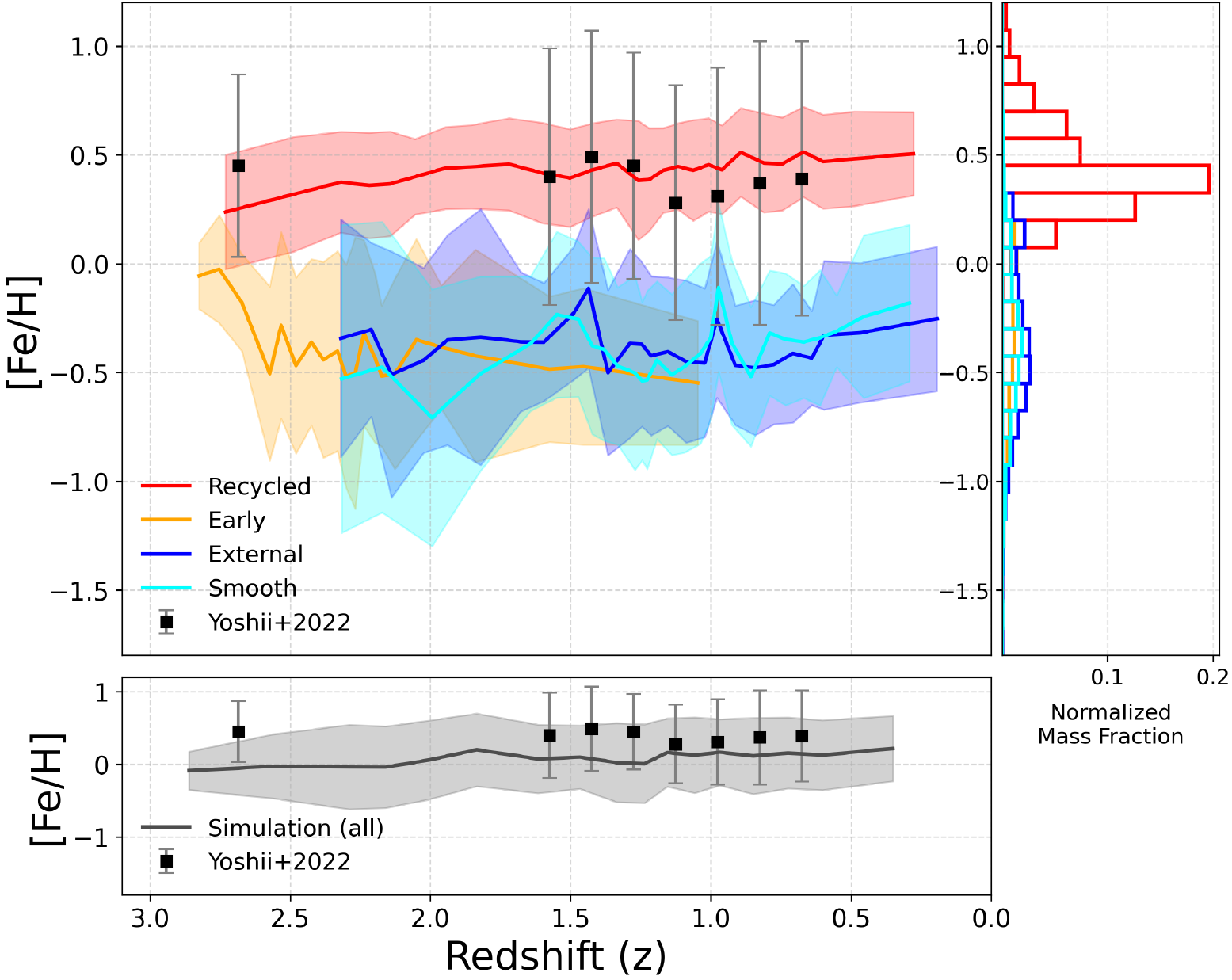}
\end{minipage}%
\begin{minipage}[t]{0.5\textwidth}
  \centering
  \includegraphics[width=\linewidth,trim=0 0 0 0,clip]{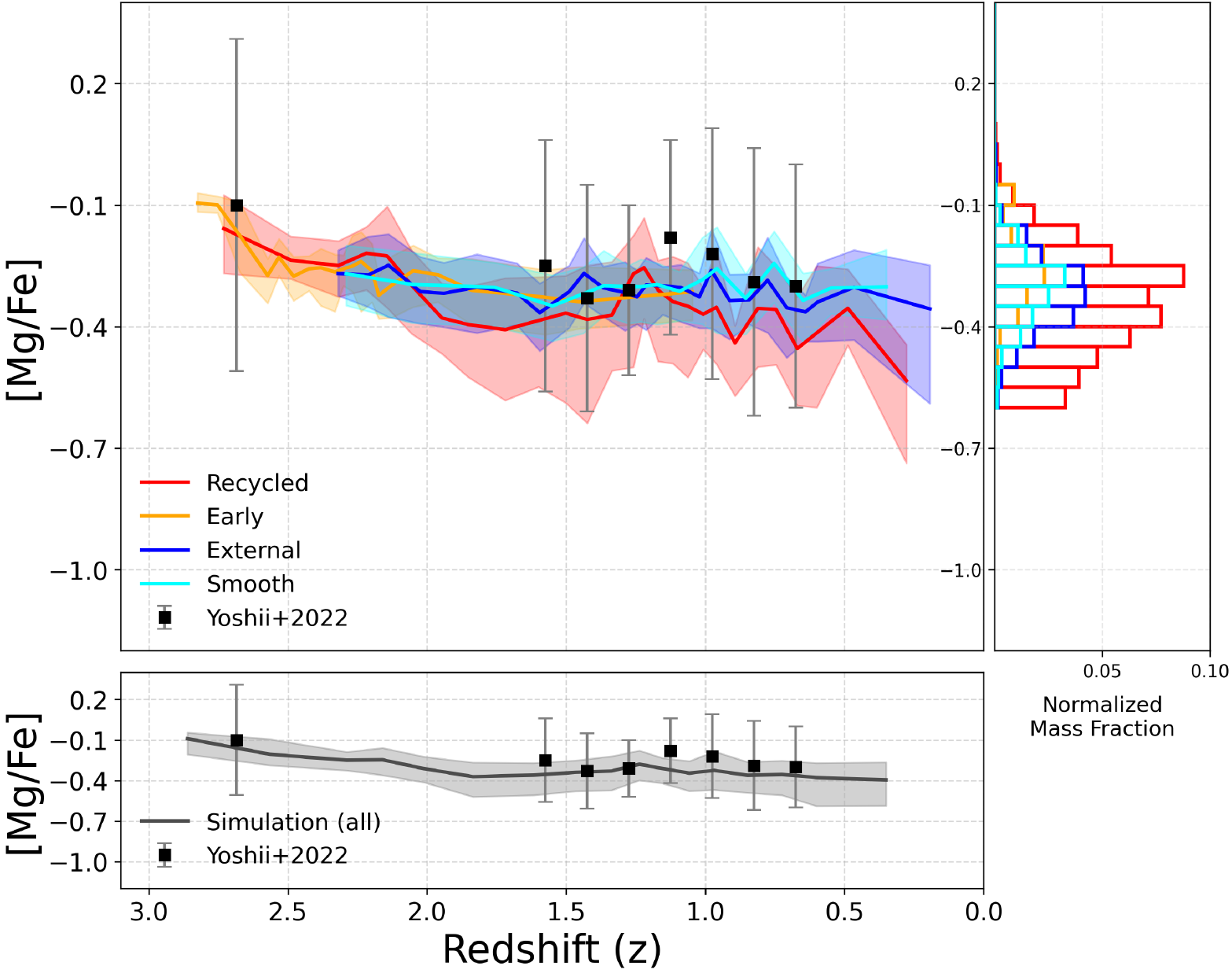}
\end{minipage}
\caption{Redshift evolution of the mass-weighted chemical abundance ratios of gas particles accreted onto the central SMBHs in the simulations. 
Shaded regions indicate the $1\sigma$ dispersion around the mass-weighted mean trends. 
{\it (Top)} Evolution of $\mathrm{[Fe/H]}$ and $\mathrm{[Mg/Fe]}$ as a function of redshift, shown separately for different gas-origin categories. The side histograms show the distributions of $\mathrm{[Fe/H]}$ and $\mathrm{[Mg/Fe]}$ for each category.
{\it (Bottom)} Mean $\mathrm{[Fe/H]}$ and $\mathrm{[Mg/Fe]}$ evolution for the full population of accreted gas particles. Observationally inferred abundance ratios from the SDSS and NTT quasar samples of \citet{Yoshii2022} are overplotted for comparison. The simulated abundance ratios exhibit only weak redshift evolution and are broadly compatible with the observational estimates within the scatter. }
\label{fig:Chemical_comparsion}
\end{figure*}

\setlength{\tabcolsep}{7pt}
\renewcommand{\arraystretch}{1.2}

\begin{table}[t]
\centering
\caption{
Spearman rank-order correlation between accretion redshift and abundance ratios of accreted gas particles. $r_s$ is the Spearman correlation coefficient and $p$ is the two-sided $p$-value for the null hypothesis of no correlation. $N_s$ denotes the number of accreted particles in each category.}
\label{tab:spearman}

\begin{tabular}{lccccc}
\hline\hline
 & \multicolumn{2}{c}{[Fe/H]} & \multicolumn{2}{c}{[Mg/Fe]} & \\[4pt]
\cline{2-3} \cline{4-5}   
Category & $r_s$ & $p$ & $r_s$ & $p$ & $N_s$ \\[4pt]
\hline
Early    & $0.25$ & $10^{-11}$ & $0.50$ & $10^{-42}$ & 630 \\
Recycled & $-0.11$  & $10^{-17}$  & $0.38$ & $10^{-153}$ & 5315 \\
External & $0.04$  & $4\times10^{-1}$ & $0.13$ & $10^{-6}$ & 1269 \\
Smooth   & $-0.07$ & $2\times10^{-2}$  & $0.03$ & $10^{-8}$ & 801\\
\hline
Total    & $-0.10$ & $10^{-84}$ & $0.33$ & $10^{-209}$ & 8015 \\
\hline

\end{tabular}

\raggedright
\end{table}

\begin{figure}
\centering
\includegraphics[width=\columnwidth]{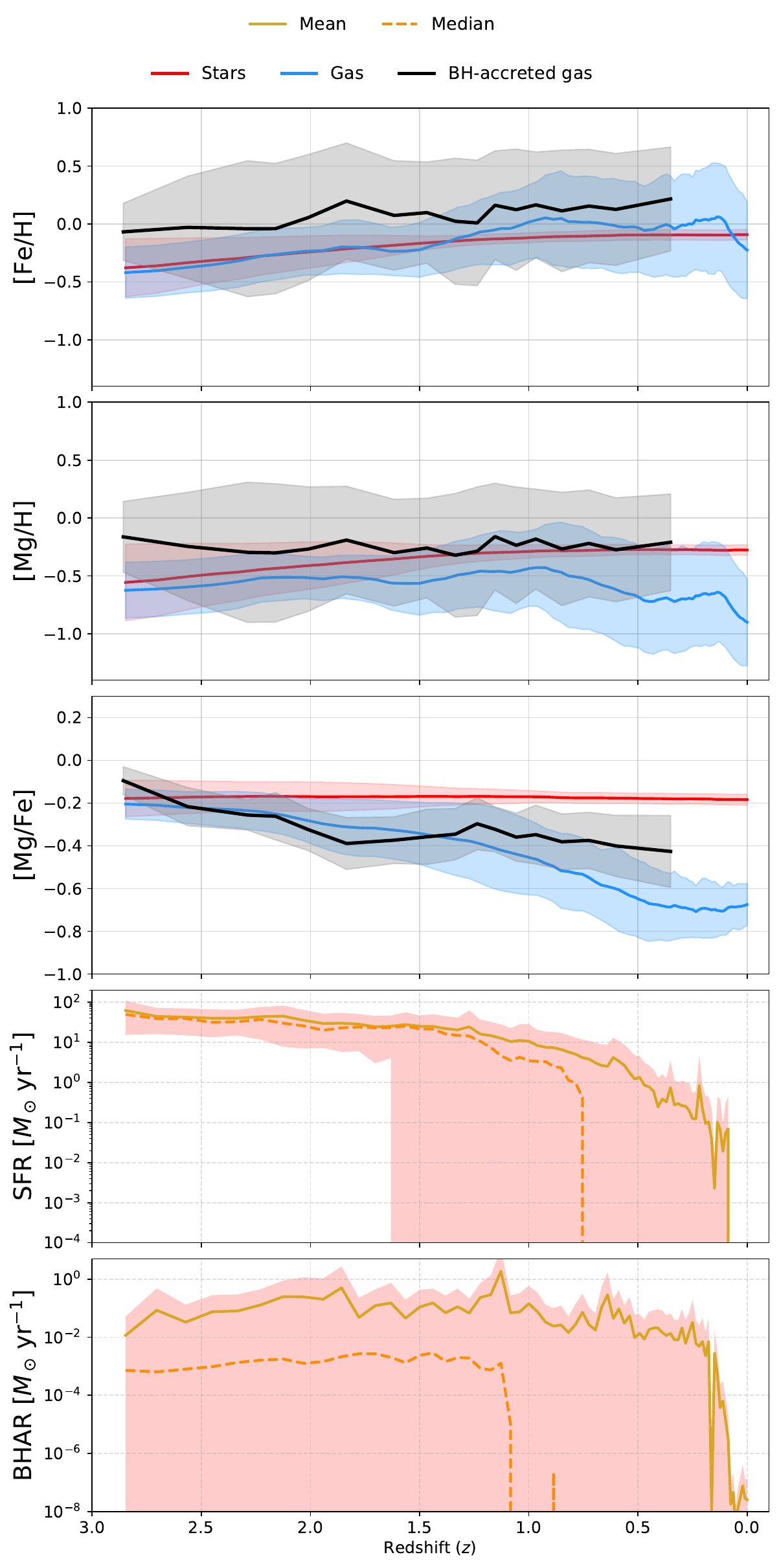}

\caption{Comparison of the redshift evolution of the star formation rate (SFR), black hole accretion rate (BHAR), and chemical abundance ratios for gas and stars within galaxies 
($r<r_{10}=0.1 \times r_{\rm vir}$), together with those for gas particles accreted onto the central SMBHs. In the SFR and BHAR panels, the solid and dashed lines indicate the mean and median values across the simulated galaxies, respectively, while the shaded regions show the corresponding $1\sigma$ scatter.  For the galaxy gas and stellar components, the mean abundance ratios and $1\sigma$ scatter are computed across all simulated galaxies at each snapshot.  For the BH-accreted gas, the abundance ratios are calculated from the full accreted-gas population shown as ``Simulation (all)'' in the bottom panels of \autoref{fig:Chemical_comparsion}. 
}

\label{fig:comparsion_galaxy}
\end{figure} 

In \autoref{fig:Chemical_comparsion}, we show the redshift evolution of the chemical abundance ratios $\mathrm{[Fe/H]}$ and $\mathrm{[Mg/Fe]}$ for gas particles accreted onto the central SMBHs, shown both separately for each gas-origin category and for the full accreted-gas population. For comparison, observationally inferred abundance ratios from the SDSS and NTT quasar samples of \citet{Yoshii2022} are also shown in both the top and bottom panels.\footnote{The observational abundance ratios from \citet{Yoshii2022} are inferred from BLR emission-line diagnostics after correcting for non-abundance effects such as Eddington-ratio and Baldwin-effect dependencies using photoionization modeling. Therefore, the comparison presented here should be interpreted qualitatively rather than as a direct one-to-one comparison of intrinsic abundances.}

The origin-separated top panels show that the accreted gas exhibits systematically different abundance patterns depending on its origin. Recycled gas is the most Fe-rich on average, consistent with its formation in the centrally enriched environment shaped by sustained stellar mass loss and metal recycling. In contrast, the Early and External components tend to be more Fe-poor on average, reflecting their earlier infall and/or chemical pre-processing in less enriched reservoirs prior to reaching the central region, although the scatter is substantial. The $\mathrm{[Mg/Fe]}$ behavior is complementary: recycled gas exhibits the lowest $\mathrm{[Mg/Fe]}$, consistent with a larger relative contribution from Type~Ia SN enrichment in the long-lived central stellar population, whereas the other components span higher values with broader dispersion.

The bottom panels show the redshift evolution of the mean abundance ratios for the full population of accreted gas particles. Overall, $\mathrm{[Fe/H]}$ exhibits only weak evolution with redshift despite substantial scatter, remaining broadly consistent over cosmic time. In contrast, $\mathrm{[Mg/Fe]}$ shows a mild systematic decline toward lower redshift. 

To quantify these evolutionary trends, we additionally perform Spearman rank-order correlation tests for each origin category, with the results summarized in \autoref{tab:spearman}. The trend is strongest for the Early gas component, which exhibits the largest monotonic correlation with accretion time ($r_{\rm s}=0.50$), while the other components show weaker or modest correlations ($r_s=0.38$ for Recycled and $r_s\leq0.13$ for External and Smooth). The declining $\mathrm{[Mg/Fe]}$ trend reflects the increasing contribution of iron enrichment from Type~Ia supernovae at later cosmic times as gas interacting with the main halo becomes progressively enriched by older stellar populations. We emphasize, however, that the inferred correlations are statistically significant primarily because of the large number of traced particles, and that the underlying evolutionary trends remain intrinsically weak.

In \autoref{fig:comparsion_galaxy}, we compare the redshift evolution of chemical abundance ratios for gas and stars within the galaxies ($r<r_{10}$) with those of gas particles accreted onto the central SMBHs. For each snapshot, the galaxy gas and stellar abundances are averaged across all simulated systems and compared with the corresponding mean abundance ratios of the accreted-gas population. This comparison allows us to assess whether the chemical properties of BH-accreted gas simply trace the bulk chemical evolution of their host galaxies.

The accreted gas exhibits systematically higher $\mathrm{[Fe/H]}$ than the ambient galaxy gas, particularly at early epochs ($z\sim1$--3), although with substantial scatter. This behavior arises because the accreted-gas population contains a significant contribution from the metal-rich recycled component, whereas the galaxy gas includes a broader mixture of chemically less enriched material. The elevated $\mathrm{[Fe/H]}$ therefore indicates that gas reaching the SMBH preferentially originates from centrally enriched regions shaped by sustained stellar evolution and metal recycling.


In $\mathrm{[Mg/Fe]}$, all components show a decreasing trend toward lower redshift, reflecting the growing contribution of delayed iron enrichment from Type~Ia supernovae. 
However, the stellar component remains systematically higher in $\mathrm{[Mg/Fe]}$ and evolves only weakly, because much of the stellar mass formed at early times and preserves the abundance pattern of the gas from which it formed. 
In contrast, the galaxy gas reaches the lowest $\mathrm{[Mg/Fe]}$ values at late times, as it continues to receive and mix with Fe-enriched material from older stellar populations. 
This component also shows an approximately constant $\mathrm{[Mg/H]}$ over the redshift range considered. 
Although the SFR decreases weakly at $z>1$ and then declines rapidly at $z<1$, the $\mathrm{[Mg/H]}$ of the galaxy gas shows a mild decreasing trend with a large scatter. 
Together with the decreasing $\mathrm{[Mg/Fe]}$ trend, this suggests that the chemical evolution of the galaxy gas is mainly associated with delayed Fe enrichment rather than a significant decline in Mg abundance. 
The BH-accreted gas also exhibits an overall nearly constant $\mathrm{[Mg/H]}$ trend, while its $\mathrm{[Fe/H]}$ increases weakly toward low redshift, indicating that its decreasing $\mathrm{[Mg/Fe]}$ is likewise primarily driven by delayed Fe enrichment.

Overall, these results indicate that gas accreted onto SMBHs does not simply mirror either the stellar or gaseous component of the host galaxy. 
Instead, its chemical evolution reflects a selective sampling of centrally enriched material, shaped by both recycled stellar ejecta and subsequent mixing with the surrounding gas reservoir. This interpretation is also suggested by the comparison between the SFR and BHAR trends in \autoref{fig:comparsion_galaxy}, which shows that black hole accretion history does not simply follow the global star formation history.
\
\section{Discussion}
\label{sec:discussion}
\subsection{Cosmological origin of metal-rich gas reaching SMBHs}
\label{subsec:cosmo_origin}
A key result of this study is that gas reaching SMBHs is already metal rich as a natural outcome of cosmological galaxy evolution. By tracing the full histories of gas particles accreted onto the central BHs, we show that their chemical properties are primarily set by large-scale processes, including star formation, stellar mass loss, and gas recycling within galaxies.

The mass census in \autoref{fig:census} demonstrates that gas originating from stellar mass loss within the main galaxy (the \emph{Recycled} component) dominates the accreted gas budget across the simulated galaxy sample. By definition, recycled gas forms in the inner regions of the galaxy ($r<r_{10}$), where sustained star formation and repeated stellar feedback rapidly enrich the surrounding medium. As a result, this component is intrinsically metal rich and reaches high $\mathrm{[Fe/H]}$ values already at early cosmic times.

Observational studies also suggest that recycled gas from stellar mass loss can play an important role in fueling AGN activity. Observations have found unexpectedly high AGN fractions in quiescent and recently quenched galaxies, where stellar winds are likely to be the dominant gas supply \citep{Aird2018,Aird2019,Aird2022,Birchall2023}. In addition, correlations between AGN luminosity and the presence of intermediate-age or young stellar populations support the idea that recycled material from evolving stars can effectively trigger BH accretion \citep{Riffel2022,Riffel2023,Ni2023}. Although these studies do not directly constrain gas metallicity, they provide independent support for the importance of recycled gas as a fueling channel for AGN, consistent with the dominant role of recycled gas found in our simulations.

The evolutionary pathways shown in \autoref{fig:trajectory} further reveal that, for gas originating outside the central galaxy (i.e., the \emph{Early}, \emph{External}, and \emph{Smooth} components), chemical enrichment predominantly occurs during their residence in the galactic environment prior to being accreted. Although these components follow distinct dynamical routes, their metallicities typically increase through cumulative enrichment via mixing with metal-enriched outflows and ambient halo gas. Consequently, while recycled gas attains high metallicity through rapid, in-situ enrichment in the central galaxy, the metal content of gas reaching the BH more generally reflects cumulative enrichment over cosmological timescales.

Consistent with this picture, the mean $\mathrm{[Fe/H]}$ of the accreted gas shows only weak evolution with redshift (\autoref{fig:Chemical_comparsion}), despite substantial scatter. Once a massive galaxy has built up a metal-rich interstellar medium, subsequent gas supplied to the BH is preferentially drawn from this enriched reservoir, naturally maintaining high metallicity over cosmic time.

Finally, the comparison in \autoref{fig:comparsion_galaxy} shows that gas reaching the BH is systematically more metal rich than the average gas within the central galaxy, particularly at intermediate redshifts ($z\sim1$--3). This offset reflects preferential access to centrally enriched regions and pre-processing by stellar evolution and feedback, decoupling the chemical evolution of BH-reaching gas from the mean galaxy-wide trends.

Overall, our results demonstrate that the metal-rich nature of gas supplied to SMBHs is a natural consequence of cosmological galaxy evolution and stellar recycling. 

\subsection{Relation to observed quasar metallicities}
\label{subsec:obs_relation}
Observational studies have investigated the metallicity of quasars across a wide range of redshifts using large samples binned in cosmic time \citep{DeRosa2011,Sameshima2017,Sameshima2020,Shin2019,Shin2021,Yoshii2022}. In these works, metallicities are inferred from emission-line ratios such as $\mathrm{Fe\,II}$ and $\mathrm{Mg\,II}$, and therefore represent population-averaged properties of heterogeneous quasar samples rather than the evolutionary history of individual systems. Several studies have found that even the most distant, luminous quasars exhibit supersolar metallicities with little evidence for redshift evolution, and have interpreted this as a consequence of BH–galaxy coevolution and observational selection effects, since only already massive, chemically evolved systems are detectable as quasars at high redshift (e.g., \citealt{Juarez2009}).

By contrast, our simulations follow the time evolution of gas properties within the same set of galaxies, allowing us to trace how the chemical abundance of gas accreted onto SMBHs evolves within individual systems over cosmic time. Each simulation snapshot captures the instantaneous chemical properties of gas particles accreted onto the central BHs, enabling a direct connection between gas origin and its enrichment history. As a result, the comparison between simulations and observations should be understood as qualitative, linking intrinsic evolutionary trends to population-averaged observational constraints.

Despite these fundamental differences in methodology and sample selection, the simulated abundance ratios of gas reaching the central regions are broadly compatible with those inferred for quasar environments. In particular, both approaches indicate that the gas associated with luminous AGN is already highly metal enriched at early epochs and exhibits only weak evolution in total metallicity with redshift. This broad compatibility suggests that the high metallicities inferred from quasar spectra need not arise solely from sub-parsec enrichment mechanisms, but can instead emerge naturally from the cosmological assembly of massive galaxies and the recycling of metal-rich stellar ejecta.

We note that observational quasar samples at low redshift are biased toward luminous, high-Eddington systems, whereas the simulated BHs at $z \lesssim 1$ typically reside in massive, quenched galaxies and accrete at lower rates. However, the observational abundance estimates used for comparison explicitly correct for luminosity-dependent effects such as the Baldwin effect, and the inferred abundance trends themselves are generally weak. 
These considerations suggest that residual differences in luminosity or accretion state are unlikely to fully account for the qualitative metallicity trends discussed here.

Overall, our results provide a cosmological context for the high metallicities inferred in quasar environments, demonstrating that metal-rich gas supply to the central BH is a generic outcome of galaxy formation and stellar evolution.

\subsection{Chemical diagnostics: [Fe/H] and [Mg/Fe]}
\label{subsec:feh_primary}
Throughout this work, we use $\mathrm{[Fe/H]}$ as the primary chemical diagnostic for characterizing the enrichment state of gas reaching the central BHs. It directly traces the cumulative build-up of metals through star formation, stellar mass loss, and gas recycling, and therefore provides a robust measure of chemical enrichment in cosmological simulations.

By contrast, abundance ratios such as $\mathrm{[Mg/Fe]}$ are more sensitive to the detailed timing of enrichment channels, including the delay-time distribution of Type~Ia supernovae, recent star formation activity, and the efficiency of metal mixing. These processes are more strongly affected by sub-resolution modeling and local physical conditions, making the interpretation of $\mathrm{[Mg/Fe]}$ inherently less robust at cosmological resolution. 

From an observational perspective, $\mathrm{[Mg/Fe]}$ is typically inferred from the $\mathrm{Mg\,II/Fe\,II}$ flux ratio, which has been widely used as a first-order proxy because the abundance ratio itself cannot be measured directly. However, the $\mathrm{Mg\,II/Fe\,II}$ ratio is not determined solely by the intrinsic chemical abundance, but is also influenced by observational biases and physical conditions in the BLR, such as gas density, ionization state, and radiative transfer effects \citep{Dong2011,DeRosa2011}. To mitigate these non-abundance effects, several studies have applied corrections for the Eddington ratio and the Baldwin effect using CLOUDY photoionization modeling, thereby converting the observed flux ratios into inferred $\mathrm{[Mg/Fe]}$ values \citep{Sameshima2017,Sameshima2020,Yoshii2022}.

Within these uncertainties, our simulations show broadly compatible $\mathrm{[Mg/Fe]}$ values and a weak decreasing trend toward lower redshift (\autoref{fig:Chemical_comparsion}, \autoref{tab:spearman}). This mild evolution can be naturally interpreted as the increasing contribution of iron from Type~Ia supernovae associated with older stellar populations at later times. Given the combined observational and theoretical uncertainties, we therefore treat $\mathrm{[Mg/Fe]}$ as a secondary diagnostic, using it to provide complementary insight into enrichment timescales rather than as a primary constraint.

\subsection{Scope and limitations of the present study}
\label{subsec:scope_limit}
As with all cosmological simulations, this study is limited by finite spatial and mass resolution and does not resolve the sub-parsec-scale BLR or accretion disk. Our results should therefore not be interpreted as a direct prediction of BLR gas properties or line-emitting conditions. Instead, they characterize the cosmological origin and chemical enrichment of the gas reservoir delivered to the smallest resolved scales around SMBHs.

Physical processes below the resolution limit may alter the instantaneous accretion rate, angular-momentum transport, and gas phase structure. However, they are less likely to erase the large-scale origin and enrichment history of the gas delivered to the central region. Recent high-resolution and torque-limited accretion studies have shown that small-scale physics can modify the accretion budget and variability of BH growth \citep[e.g.][]{Angles-Alcazar2017a,Angles-Alcazar2021}. Within this scope, our simulations provide cosmological boundary conditions for metal-rich gas supply to SMBHs, rather than a direct model of BLR-scale enrichment.

In addition, our sample focuses on massive group-scale galaxies at low redshift, and the relative importance of fueling channels may differ in lower-mass or gas-rich systems.

\section{Summary}
\label{sec:summary}
We investigate the chemical properties of gas supplied to SMBHs using cosmological zoom-in simulations, building on the particle-tracking methodology introduced in \citetalias{Choi2024}. By tracing the full enrichment histories of gas particles accreted onto the central BHs, we focus on the smallest resolved gas reservoir that reaches the BH, with the aim of characterizing the cosmological origin and chemical pre-processing of BH fueling gas.

Using a sample of 30 massive galaxies hosting SMBHs, we classify accreted gas into four origin categories: Recycled, Early, External, and Smooth, and examine how their distinct dynamical pathways shape their chemical properties. Our main findings can be summarized as follows:
\begin{enumerate}
    \item Gas originating from stellar mass loss within the central galaxy (the Recycled component) dominates the BH gas supply and is intrinsically metal rich from early cosmic times, highlighting the central role of stellar evolution and gas recycling in setting the chemical state of BH-reaching gas.
    
    \item Gas originating outside the central galaxy undergoes progressive chemical enrichment during its residence in the galactic environment prior to being accreted, indicating that chemical pre-processing largely precedes accretion and reflects cumulative enrichment over cosmological timescales.
    
    \item The total metallicity of accreted gas, as traced by $\mathrm{[Fe/H]}$, exhibits only weak evolution with redshift, despite substantial scatter. This behavior arises naturally once massive galaxies establish metal-rich central gas reservoirs, from which subsequent BH fueling gas is drawn.
    
    \item The abundance ratio $\mathrm{[Mg/Fe]}$ shows a weak decreasing trend toward lower redshift. This mild evolution is consistent with the increasing contribution of iron from Type~Ia supernovae associated with older stellar populations, and is treated as a secondary diagnostic in this work.
    
    \item By comparing the chemical properties of accreted gas with those of gas and stars in the central regions of their host galaxies, we find that BH–reaching gas is systematically more metal rich than the average galaxy gas at intermediate redshifts ($z\sim1$--3), reflecting preferential access to centrally enriched material and local chemical pre-processing.
\end{enumerate}

Overall, our results demonstrate that the metal-rich nature of gas supplied to SMBHs is driven primarily by recycled stellar ejecta and is therefore a natural outcome of cosmological galaxy evolution. By establishing the cosmological boundary conditions for chemically enriched gas reaching the central BH, this study provides a physically grounded framework for interpreting the high metallicities inferred in quasar environments, independent of unresolved accretion-scale physics.

\begin{acknowledgments}
We thank the anonymous referee for their constructive comments.
This work was supported by the National Research Foundation of Korea (NRF) grant funded by the Korea government (MSIT) (No. RS-2025-00515276) for DK and EC. This work was also supported by the 2023 Advanced Facility Fund of the University of Seoul. 
TN acknowledges the support of the Deutsche Forschungsgemeinschaft (DFG, German Research Foundation) under Germany's Excellence Strategy - EXC-2094 - 390783311 of the DFG Cluster of Excellence ``ORIGINS''. The Flatiron Institute is supported by the Simons Foundation.
\end{acknowledgments}

\bibliography{library}
\bibliographystyle{aasjournalv7}
\end{document}

%% file: defs.tex
\def\galaxiespprox{\mathrel{\vcenter{\offinterlineskip \hbox{$>$}
    \kern 0.3ex \hbox{$\sim$}}}}
\def\lapprox{\mathrel{\vcenter{\offinterlineskip \hbox{$<$}
    \kern 0.3ex \hbox{$\sim$}}}}
\newcommand{\beq}{\begin{equation}} 
\newcommand{\eeq}{\end{equation}}

\def\Sig15{\Sigma_{1.5}}

\def\kpch{{\rm\thinspace kpc} \thinspace \it h^{\rm -1}}

\def\Msunh{\hbox{$\thinspace \rm \thinspace M_{\odot}\thinspace \it h^{\rm -1}$}}

\def\r10{r_{10}}

\def\strom{Str$\rm \ddot{o}$mgren$\thinspace$}

\def\delm12{M_{12}}
\def\sigm1{\sigma(m_{1})}